\documentclass[epj,nopacs]{svjour}
\usepackage{epsfig}
\usepackage{amssymb,amsmath}

\usepackage{color}

\usepackage{lineno}
\usepackage{vruler}

\newcommand{\beq}  {\begin{equation}}
\newcommand{\eeq}  {\end{equation}}
\newcommand{\beqn}  {\begin{eqnarray}}
\newcommand{\eeqn}  {\end{eqnarray}}

\newcommand{\desy}{DESY}

\def\nowat[#1]#2{\(^,\)\footnote[#1]{#2}}

\newcommand{\etal}{\textit{et~al.}}
\newcommand{\JOURNAL}[4]{#1 {\bf #2}, #3 (#4)}

\newcommand{\PRL}[3]{\JOURNAL{Phys. Rev. Lett.}{#1}{#2}{#3}}

\newcommand{\PRD}[3]{\JOURNAL{Phys. Rev. D}{#1}{#2}{#3}}
\newcommand{\PLB}[3]{\JOURNAL{Phys. Lett. B}{#1}{#2}{#3}}
\newcommand{\NIMA}[3]{\JOURNAL{Nucl. Instrum. Methods A}{#1}{#2}{#3}}

\newcommand{\NPB}[3]{\JOURNAL{Nucl. Phys. B}{#1}{#2}{#3}}

\newcommand{\PR}[3]{\JOURNAL{Phys. Rept.}{#1}{#2}{#3}}

\newcommand{\ANN}[3]{\JOURNAL{Ann. Phys.}{#1}{#2}{#3}}
\newcommand{\JHEP}[3]{\JOURNAL{JHEP}{#1}{#2}{#3}}

\begin{document}

\hugehead


\title{Measurement of the virtual-photon asymmetry $\mathbf{A_2}$ and
the spin-structure function $\mathbf{g_{\hspace*{.03cm}2}}$ of the proton
}

\author{{\bf 
The HERMES Collaboration \medskip \\
A.~Airapetian$^{12,15}$,
N.~Akopov$^{26}$,
Z.~Akopov$^{5}$,
E.C.~Aschenauer$^{6}$\nowat[1]{Now at: Brookhaven National Laboratory, Upton,
New York 11772-5000, USA},
W.~Augustyniak$^{25}$,
R.~Avakian$^{26}$,
A.~Avetissian$^{26}$,
E.~Avetisyan$^{5}$,
S.~Belostotski$^{18}$,
N.~Bianchi$^{10}$
% ...
}}

\institute{ 
$^1$Physics Division, Argonne National Laboratory, Argonne, Illinois 60439-4843, USA\\
$^2$Istituto Nazionale di Fisica Nucleare, Sezione di Bari, 70124 Bari, Italy\\
$^3$School of Physics, Peking University, Beijing 100871, China\\
$^4$Nuclear Physics Laboratory, University of Colorado, Boulder, Colorado 80309-0390, USA\\
$^5$DESY, 22603 Hamburg, Germany\\
$^6$DESY, 15738 Zeuthen, Germany\\
$^7$Joint Institute for Nuclear Research, 141980 Dubna, Russia\\
$^8$Physikalisches Institut, Universit\"at Erlangen-N\"urnberg, 91058 Erlangen, Germany\\
$^9$Istituto Nazionale di Fisica Nucleare, Sezione di Ferrara and Dipartimento di Fisica, Universit\`a di Ferrara, 44100 Ferrara, Italy\\
$^{10}$Istituto Nazionale di Fisica Nucleare, Laboratori Nazionali di Frascati, 00044 Frascati, Italy\\
$^{11}$Department of Physics and Astronomy, Ghent University, 9000 Gent, Belgium\\
$^{12}$Physikalisches Institut, Universit\"at Gie{\ss}en, 35392 Gie{\ss}en, Germany\\
$^{13}$SUPA, School of Physics and Astronomy, University of Glasgow, Glasgow G12 8QQ, United Kingdom\\
$^{14}$Department of Physics, University of Illinois, Urbana, Illinois 61801-3080, USA\\
$^{15}$Randall Laboratory of Physics, University of Michigan, Ann Arbor, Michigan 48109-1040, USA \\
$^{16}$Lebedev Physical Institute, 117924 Moscow, Russia\\
$^{17}$National Institute for Subatomic Physics (Nikhef), 1009 DB Amsterdam, The Netherlands\\
$^{18}$Petersburg Nuclear Physics Institute, Gatchina, 188300 Leningrad Region, Russia\\
$^{19}$Institute for High Energy Physics, Protvino, 142281 Moscow Region, Russia\\
$^{20}$Institut f\"ur Theoretische Physik, Universit\"at Regensburg, 93040 Regensburg, Germany\\
$^{21}$Istituto Nazionale di Fisica Nucleare, Sezione di Roma, Gruppo Collegato Sanit\`a and Istituto Superiore di Sanit\`a, 00161 Roma, Italy\\
$^{22}$TRIUMF, Vancouver, British Columbia V6T 2A3, Canada\\
$^{23}$Department of Physics, Tokyo Institute of Technology, Tokyo 152, Japan\\
$^{24}$Department of Physics and Astronomy, VU University, 1081 HV Amsterdam, The Netherlands\\
$^{25}$National Centre for Nuclear Research, 00-689 Warsaw, Poland\\
$^{26}$Yerevan Physics Institute, 375036 Yerevan, Armenia\\
}

\date{Received: \today / Revised version:}

\abstract{
A measurement of the virtual-photon asymmetry $A_2(x,Q^2)$ and of the
spin-structure function $g_2(x,Q^2)$ of the proton 
are presented for the kinematic range
0.004 $<$ $x$ $<$ 0.9 and 
0.18 GeV$^2$ $<$ $Q^2$ $<$ 20 GeV$^2$.
The data were collected by the
HERMES experiment at the HERA storage ring at DESY 
while studying inclusive deep-inelastic scattering of 27.6 GeV
longitudinally polarized leptons  off a transversely polarized hydrogen gas target.
The results are consistent
with previous experimental data from CERN and SLAC.
For the $x$-range covered, the measured integral of $g_2(x)$ converges to
the null result of the Burkhardt--Cottingham sum rule.  
The $x^2$ moment of the twist-3 contribution to $g_2(x)$
is found to be compatible with zero.
}

\maketitle

The description of inclusive deep-inelastic scattering of longitudinally polarized 
charged leptons off polarized nucleons requires two 
nucleon spin-structure functions, $g_1(x,Q^2)$ and $g_2(x,Q^2)$, in addition to the well-known
structure functions $F_1(x,Q^2)$ and $F_2(x,Q^2)$ \cite{anselmino}.
Here, $ - Q^2$ is the squared four-momentum of the exchanged virtual photon
with laboratory energy $\nu$, $x=Q^2/(2M\nu)$ is the Bjorken scaling variable, 
and $M$ is the nucleon mass. 
In the quark-parton model (QPM), the spin structure function $g_1(x,Q^2)$
can be interpreted as a charge-weighted sum of the quark helicity distributions 
$\Delta q(x,Q^2)$ describing a longitudinally polarized nucleon, 
\begin{equation}
g_1(x,Q^2) = \frac{1}{2} \sum_q e_q^2 \Delta q(x,Q^2).
\end{equation}

The spin structure function $g_2(x,Q^2)$ does not have such a probabilistic
interpretation in the QPM.
Its properties can be interpreted 
in the framework of the operator product expansion (OPE) analysis \cite{shuryak,Jaf90,jaffe-ji91}, 
which shows that $g_2(x, Q^2)$ is related to the matrix elements of both twist-2 and 
twist-3 operators.
Neglecting quark mass effects,  $g_2(x,Q^2)$ can be written as a sum of two terms
\beq
\label{eq:g2sum}
g_2(x, Q^2) = g_2^{\rm WW}(x, Q^2) + \bar{g}_2(x, Q^2) \, .
\eeq
Here, $g_2^{\rm WW}(x, Q^2)$ is the twist-2 part derived
by Wandzura and Wilczek \cite{Wan77}
\beq
\label{eq:g2ww}
 g_2^{\rm WW}(x, Q^2) = -g_1(x, Q^2) + \int_x^1 g_1(y, Q^2) \frac{d\, {y}}{y} \, .
\eeq
The second term in Eq.~(\ref{eq:g2sum}), $\bar{g}_2(x, Q^2)$, is the twist-3 part of $g_2(x,Q^2)$. 
It arises from quark-gluon correlations in the nucleon and is the most interesting part
of the function. The $x^2$ moment of $\bar{g}_2(x, Q^2)$, 
\beq
\label{eq:d2mom}
d_2(Q^2) = 3 \int_0^1  x^2 \, \bar{g}_{2}(x, Q^2) \, dx \, , 
\eeq
can be calculated on the lattice (see, \textit{e.g.}, \cite{Gockeler,Gockeler2}, where $d_2$ 
is defined with an additional factor of two with respect to (\ref{eq:d2mom})).
The moment $d_2$ has also been linked to the transverse force acting on
the quark that absorbed the virtual photon in a transversely polarized
nucleon, and thus to the Sivers effect \cite{sivers,burkardt,hermessivers}.

The Burkhardt--Cottingham sum rule \cite{bcsumrule} for $g_2$ at large $Q^2$, 
\beq
\label{eq:bcsumrule}
  \int_0^1 g_2(x, Q^2) \, dx = 0 \, ,
\eeq
does not follow from the OPE. Its validity relies on an assumed Regge behaviour of $g_2$ 
at low $x$. In the absence of higher twist contributions to the function  $g_2$, 
{\it i.e.}, $\bar{g}_2(x) \equiv 0$, 
the sum rule would automatically be fulfilled.
Hence a violation of the sum rule would indicate the presence of higher-twist 
contributions. 

The spin structure functions $g_1(x,Q^2)$ and $g_2(x,Q^2)$ can be related to the
virtual photon-ab\-sorp\-tion asymmetries $A_1(x,Q^2)$ and $A_2(x,Q^2)$ \cite{anselmino}
\beqn
\label{eq:eqa1}
A_1 & = & \frac{\sigma^{\cal{T}}_{1/2} - \sigma^{\cal{T}}_{3/2}}
{\sigma^{\cal{T}}_{1/2} + \sigma^{\cal{T}}_{3/2}}
 =
\frac{g_1 - \gamma^2 g_2}{F_1} \, , \\
\label{eq:eqa2}
A_2 & = & \frac{2 \sigma^{\cal{LT}}}{\sigma^{\cal{T}}_{1/2} + \sigma^{\cal{T}}_{3/2}} = 
\gamma \, \frac{g_1 + g_2}{F_1} \, .
\eeqn
Here, $\sigma^{\cal{T}}_{1/2}$ and $\sigma^{\cal{T}}_{3/2}$ are 
the transverse virtual-photon ab\-sorp\-ti\-on 
cross sections for total photon plus nucleon angular momentum projection on the photon
direction of $1/2$ and $3/2$, respectively. The cross-section $\sigma^{\cal{LT}}$ 
arises from the interference between the transverse and longitudinal photon-nucleon 
amplitudes, with
{\normalsize $\gamma = 2 M x/\sqrt{Q^2}$}. All of the $\sigma$'s are differential
cross sections depending on $x$ and $Q^2$, but this dependence was omitted for brevity.

The measurement of the structure function $g_2$ requires a longitudinally
polarized beam and a transversely polarized target. In this case, the inclusive differential
cross section can be represented as a sum of two terms, 
the po\-la\-ri\-za\-tion-averaged
part, $\sigma_{UU}$, and the po\-la\-ri\-za\-tion-related part, $\sigma_{LT}$.
Here, the subscript {\scriptsize $UU$} indicates that both the beam and the target 
are  unpolarized,
while the subscript {\scriptsize $LT$} indicates a longitudinally polarized beam and 
a transversely polarized target.
The polarization-related part of the cross section at Born level, 
{\it i.e.}, in the one-photon approximation, is given by \cite{Jaf90}
\beqn
\label{eq:sigpol}
\frac{{d}^3 \sigma_{LT}}{{d}x {d}y {d}\phi} \, &=& \,  
- h_l  \cos\phi  \frac{4 \alpha^2}{Q^2} 
\gamma  \sqrt{1 - y - \frac{\gamma^2 y^2}{4}} \nonumber \\
 & \times &
       \Bigl ( \frac{y}{2} g_1(x,Q^2) + g_2(x,Q^2) \Bigr ) \, .
\eeqn
Here, $h_l = +1$ ($-1$) for a lepton beam with positive (negative) helicity, 
$\alpha$ is the fine-structure constant, and
$y = \nu/E$, where $E$ is the incident lepton energy. 
The angle $\phi$ is the azimuthal angle about the beam direction between
the lepton scattering plane and the ``upwards'' target spin direction.
The polarization-related cross section $\sigma_{LT}$ is significantly smaller
than the po\-la\-ri\-za\-ti\-on-ave\-ra\-ged part $\sigma_{UU}$
and therefore its measurement requires high statistical precision.
Up to now, the function $g_2$ and the asymmetry $A_2$ have been extracted 
\cite{smc97,e143,e155} to less accuracy than $g_1$ and $A_1$.

A measurement of the inclusive cross sections (\ref{eq:sigpol}) at angles $\phi$ 
and $\phi+\pi$ allows one to construct the asymmetry $A_{LT}$,
\beqn
\label{eq:asysig}
A_{LT}(x,Q^2,\phi) =  
h_l \, \frac{\sigma(x,Q^2,\phi) - \sigma(x,Q^2,\phi+\pi)}
{\sigma(x,Q^2,\phi) + \sigma(x,Q^2,\phi+\pi)} \nonumber \\
= 
h_l \, \frac{\sigma_{LT}(x,Q^2,\phi)}{\sigma_{UU}(x,Q^2,\phi)} =
-  A_{T}(x,Q^2) \,\cos\phi \, , 
\eeqn
which defines the asymmetry amplitude $A_{T}(x,Q^2)$. 
This amplitude contains all information on the function $g_2$ and the asymmetry $A_2$.
Their extraction requires the knowledge of $\sigma_{UU}(x, Q^2)$, which can be 
expressed by the structure functions $F_{1, 2}(x, Q^2)$ or,
equivalently, parameterizations of the function $F_2(x, Q^2)$ and the ratio of 
longitudinal to transverse virtual-photon  absorption cross sections $R = R( x, Q^2)$.
The extraction of the structure function $g_2(x,Q^2)$ from the asymmetry amplitude $A_T$ is
analogous to the extraction of $g_1(x,Q^2)$ from the longitudinal asymmetry as
described in \cite{HERM07}.
The function $g_2$ can be extracted from the measured asymmetry amplitude $A_T$ and
parameterizations of previous measurements of $\sigma_{UU}$ and $g_1$, using (\ref{eq:sigpol})
and (\ref{eq:asysig}). Also $F_1$ can be computed from parameterizations of $F_2$ and $R$.
This leads with (\ref{eq:eqa2}) to the following relations
\begin{equation}\label{eq:g2}
g_2= \frac{F_1}{\gamma (1+\gamma\xi)} \left( \frac{A_{T}}{d} \,-\,
     (\gamma \, - \,  \xi) \frac{g_1}{F_1}\right) \,,
\end{equation} 
\begin{equation}\label{eq:A2}
A_2 = \frac{1} {1+\gamma\xi} \left( \frac{A_{T}}{d} \, + \, 
      \xi(1+\gamma^2) \frac{g_1} {F_1} \right) \, ,
\end{equation} 
with 
\beqn
d &=& \frac{\sqrt{1-y-\gamma^2 y^2/4}}{(1-y/2)} D \,,  \\
\xi &=& \frac{\gamma (1-y/2)}{(1+\gamma^2y/2)} \,,   
\eeqn
\begin{equation}
D =  \frac{y(2-y) (1+\gamma^2 y/2)}{y^2(1+\gamma^2)
  + 2(1-y-\gamma^2 y^2/4)(1+R)} \, .
\end{equation}
However, it is not obvious from these relations that the extraction of $g_2$ is independent 
of correlated variations in values of $F_1$, $F_2$ and $R$ that conserve the directly
measured values of $\sigma_{UU}$.

This paper reports a new measurement of the function $g_2$ and the asymmetry $A_2$.
The data were collected during the years 2003 -- 2005 
with the HERMES spectrometer \cite{hermes} 
using a longitudinally polarized positron or electron beam of energy 27.6 GeV scattered off 
a transversely  polarized \mbox{target \cite{target}} of pure hydrogen gas 
internal to the HERA lepton storage ring at DESY. 
The usage of a pure target avoids the complications of nuclear corrections 
present in previous measurements.
The open-ended target cell was fed by an atomic-beam source \cite{abs2003} 
based on Stern--Gerlach
separation combined with radio-frequency transitions 
between hydrogen hyperfine states. The nuclear polarization of the atoms was flipped
at 1--3 minute time intervals, while both the polarization magnitude 
and the atomic fraction inside the target cell were continuously measured \cite{polarmeas}. 
The average magnitude of the proton polarization was $0.78 \pm 0.04$. 
The lepton beam (positrons during 2003 -- 2004 
and electrons in 2005) was self-po\-la\-riz\-ed in the transverse direction 
due to the asymmetry in the emission of synchrotron radiation \cite{sok}
in the arcs of the HERA storage ring.
Longitudinal orientation of the beam polarization was obtained 
by using a pair of spin rotators \cite{buon} located before and
after the interaction region of the HERMES spectrometer.
The sign of the beam polarization was reversed every few months. 
The beam polarization was measured by two independent
HERA polarimeters \cite{barber,tpol,lpol}.
The average magnitude of the beam polarization was found to be $0.34 \pm 0.01$. 
The scattered leptons were detected by the HERMES spectrometer
within an angular acceptance of $\pm 170$ mrad horizontally  
and $\pm(40 - 140)$ mrad vertically.
The leptons were identified using the information from an electromagnetic 
calorimeter, 
a transition-radiation detector, a preshower scintillating counter and 
a dual-radiator ring-imaging \u Cerenkov detector. 
The identification efficiency for leptons with momentum larger than 2.5 GeV
exceeds 98\%, while the hadron contamination in the lepton sample 
is found to be less than 1\%.
The luminosity monitor \cite{lum} measured $e^+e^-$ $(e^-e^-)$ pairs from Bhabha (M\o ller) 
scattering of beam positrons (electrons) off the target gas electrons, 
and $\gamma\gamma$ pairs from $e^+e^-$ annihilation in two NaBi(WO$_4)_2$
electromagnetic calorimeters, which were mounted symmetrically on either side
of the beam line. 
Tracking corrections were applied for the deflections of the scattered particles 
caused by the vertical 0.3 T target holding field, 
with little effect on the extracted asymmetries.

Most of the details of the analysis follow the inclusive analysis described in
\cite{HERM07}.
The kinematic constraints imposed on the events were:
0.18 GeV$^2$ $<$ $Q^2$ $<$ 20 GeV$^2$, 
invariant mass of the virtual photon--nucleon system 
$W > 1.8$ $\mathrm{GeV}$,
$0.004 < x < 0.9$, and $0.10 < y < 0.91$. 
After applying data quality criteria, 10.2 $\times$ 10$^6$ events were available
for the asymmetry analysis. 
The kinematic region covered by the experiment in  ($x$, $Q^2$)-space was divided
into nine bins in $x$. Each of the seven $x$-bins in the region $x > 0.023$
was subdivided into three logarithmically equidistant bins in $Q^2$.
The range in $\phi$-space ($2 \pi$) was divided into 10 bins.
Two of the $\phi$-bins cover the shielding steel-plate region of the spectrometer 
and thus cannot be used for the analysis.
The data were corrected for the $e^+e^-$ charge-symmetric background \cite{HERM07}, 
which amounted in total to about 1.8\% of the events, reaching the largest contribution 
of about 14\% at small values of $x$.

The measurement of the asymmetry $A_{LT}(x,Q^2,\phi)$ given by (\ref{eq:asysig}) can be 
performed by either reversing
the transverse target polarization and comparing the number of events in the same part
of the detector, or by comparing the number of events in the upper and lower part 
of the detector for the same upward or downward target polarization direction.
The first method provides a better cancellation of acceptance effects and was chosen
to obtain the asymmetry
\begin{flalign}
\label{eq:asym}
& A_{LT}(x,Q^2,\phi,h_l) = \nonumber \\
& h_l \, \frac
{N^{h_l \Uparrow}(x,Q^2,\phi) {\cal   L}^{h_l \Downarrow}\,-
               \,N^{h_l \Downarrow}(x,Q^2,\phi) {\cal   L}^{h_l \Uparrow}}
{N^{h_l \Uparrow}(x,Q^2,\phi) {\cal L}_p^{h_l \Downarrow}\,+
\,N^{h_l \Downarrow}(x,Q^2,\phi) {\cal L}_p^{h_l\Uparrow}}\,.
\end{flalign}
Here, $N^{h_l {\Uparrow(\Downarrow)}}$ is the number of 
scattered leptons in one bin of the 3-dimensional space $(x, Q^2, \phi)$ 
for the case of the incident lepton with helicity $h_l$
when the direction of the proton spin points up (down). 
${\cal L}^{h_l \Uparrow(\Downarrow)}$ and 
${\cal L}^{h_l \Uparrow(\Downarrow)}_p$ are 
the corresponding integrated luminosities and the integrated luminosities 
weighted with the absolute value of the beam and target polarization product, 
respectively
\beqn
{\cal   L}^{h_l \Uparrow (\Downarrow)} & = & 
       \int \!{d}t \,L^{h_l \Uparrow (\Downarrow)}(t) \tau(t) \, ,\\
{\cal   L}_p^{h_l \Uparrow (\Downarrow)} & = & 
       \int \!{d}t \,L^{h_l \Uparrow (\Downarrow)}(t) \mid P_B(t) P_T(t) \mid \tau(t) .
\eeqn
Here, $L(t)$ is the luminosity, $\tau(t)$ is the trigger live-time factor, and $P_B$ and 
$P_T$ are the beam and target polarizations, respectively.
The asymmetries evaluated according to (\ref{eq:asym}) were found to be 
consistent for the two beam helicity states. Therefore they were combined 
in the further analysis.
Finally, the asymmetry given by (\ref{eq:asym}) was unfolded for radiative 
and instrumental smearing effects to obtain the asymmetry corresponding 
to single-photon exchange in the scattering process.
Radiative corrections were calculated using a Monte-Carlo generator \cite{Aku98}.
The unfolding procedure is analogous to that used previously in other HERMES 
analyses \cite{HERM07,HERM05,HERMb1}.
It inflates the size of the statistical uncertainties
especially in the lowest $Q^2$-bins at a given value of $x$.
The magnitude of inflation reaches almost a factor of two at low values of $x$.
The subdivision of $x$-bins in the range $x > 0.023$ into three bins in $Q^2$
decreases the error inflation by about a factor of 1.5
because at larger $Q^2$ the amount of smearing between $x$-bins is smaller
and the prefactors of $A_{T}$ in (\ref{eq:g2}) and (\ref{eq:A2}) are 
larger in magnitude. After the unfolding procedure
the central values of $g_2$ and $A_2$ changed less than the initial statistical
uncertainties.
As a consequence of the unfolding procedure, the resulting data points are no longer
correlated systematically through radiative and instrumental smearing effects, 
but are only statistically correlated \cite{HERM07}.
The procedure generates a statistical covariance matrix for the data points.

In every $(x,Q^2)$-bin the amplitude $A_{T}(x,Q^2)$ was obtained by fitting the unfolded 
asymmetries with the function $f(\phi) = -A_{T}(x,Q^2)\cos\phi$.
Finally, the asymmetry $A_2(x,Q^2)$ and the function $g_2(x,Q^2)$ 
were evaluated from the amplitude $A_{T}$ 
and the previously measured function $g_1$, for which a world-data 
parameterization \cite{Ant00} was employed, 
using (\ref{eq:g2}) and (\ref{eq:A2}).
The structure function 
\beqn
F_1(x,Q^2) = F_2(x,Q^2) (1+\gamma^2)/[2x(1+R(x, Q^2))]
\eeqn
was calculated using a parameterization
of the structure function $F_2(x,Q^2)$ \cite{Gab07} and 
the ratio $R(x,Q^2)$ \cite{Abe99}.
All kinematic factors in (\ref{eq:g2}) and (\ref{eq:A2}), 
and the functions
$F_1$ and $g_1/F_1$ were calculated at the average values of $x$ 
and $Q^2$ in each ($x$, $Q^2$)-bin after unfolding.

The uncertainties in the measurements of the beam and target polarizations 
produce in total 
a $10\%$ scale uncertainty on the value of $A_{T}$.
Other sources of systematic uncertainties such as acceptance effects,
small beam and spectrometer misalignments, 
uncertainties in the target polarization direction, 
correction for track deflection in the vertical target holding field, 
the unfolding procedure and a possible correlation between prefactors of $A_{T}$ and
$A_{T}$ itself in (\ref{eq:g2}) and (\ref{eq:A2})
were evaluated by Monte-Carlo studies.
Uncertainties stemming from   parameterizations of 
$g_1(x, Q^2)$, $F_2(x, Q^2)$, and $R(x, Q^2)$ were estimated also.
In the error propagation to $g_2$, the uncertainty in $R(x, Q^2)$ was not included 
in addition to that of $F_2(x, Q^2)$, since they are strongly correlated 
as explained in \cite{HERM07}.
The total systematic uncertainty
was evaluated as the quadratic sum of all the considered sources.
Its magnitude is less than the magnitude of the statistical uncertainty.

\begin{figure}[hbt]
\centering
\hspace*{-5mm}\includegraphics[width=10.1cm] {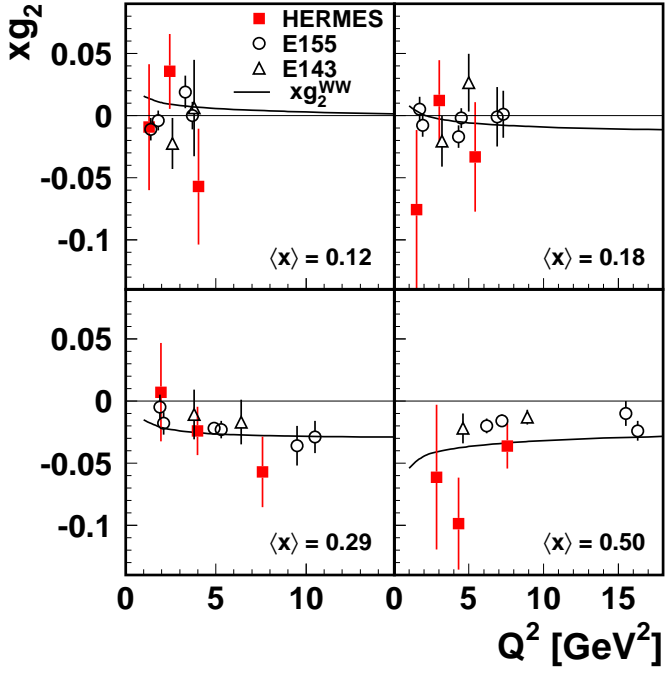}
\caption{
The spin-structure function $xg_2(x,Q^2)$ of the proton as a function of $Q^2$ for 
selected values of $x$. Data from the experiments
E155 \cite{e155} and E143 \cite{e143} are presented also. The average values of $x$
for these two experiments are slightly different from the HERMES values of 
$\langle x \rangle$ indicated in the panels. 
The error bars represent the quadratic sum of the statistical and systematic uncertainties. 
The solid curve is the result of the Wandzura--Wilczek relation 
(\ref{eq:g2ww})
}\label{figQ2}
\end{figure}

Figure~\ref{figQ2} shows the values of $xg_2$ as a function of $Q^2$ 
for the bins with $x>0.1$, which have sufficient coverage in $Q^2$,
along with results from the E143~\cite{e143} and E155~\cite{e155} experiments at SLAC.
The entire set of measured data and average values of $x$ and $Q^2$ are presented in 
Table~\ref{tab:bigtab}.
Within the accuracy of the data, they are in agreement with the other experiments. 
Also shown is the Wandzura--Wilczek term $g_2^{\rm WW}$, 
which was evaluated according to (\ref{eq:g2ww}). 
A world data parameterization of $g_1(x, Q^2)$ \cite{Ant00} was used for the calculation.

In order to study the $x$ dependence, $A_2(x,Q^2)$ and $g_2(x,Q^2)$ 
in bins covering the same $x$ range but with different $Q^2$ values 
were evolved to their mean value of $Q^2$ and then averaged.
The evolution of $A_2(x,Q^2)$ was carried out assuming 
that the product $\sqrt{Q^2} A_2$ does not depend on $Q^2$, which follows 
from (\ref{eq:eqa2}),
since $g_1/F_1$ is known to vary only weakly over $Q^2$.
The structure function $g_2(x,Q^2)$ was evolved assuming that its $Q^2$ dependence is
analogous to that for the Wandzura-Wilczek part of $g_2$.

The averaged results for $xg_2$ and $A_2$ and the statistical and systematic 
uncertainties are listed in Table~\ref{tab:tab1}, where the average values 
of $x$ and $Q^2$ are also given.
The quoted statistical uncertainties correspond
to the diagonal elements of the covariance matrix obtained
from the unfolding algorithm. The correlation matrix for $xg_2$ in nine $x$-bins is presented in
Table~\ref{tab:tab2}.\footnote{It is also available in 23 bins for the data in Table~\ref{tab:bigtab}
at http://inspirehep.net/record/1082840 or from management@hermes.desy.de.}

\begin{figure}[hbtp]
\centering
\hspace*{-5mm}\includegraphics[width=10.1cm] {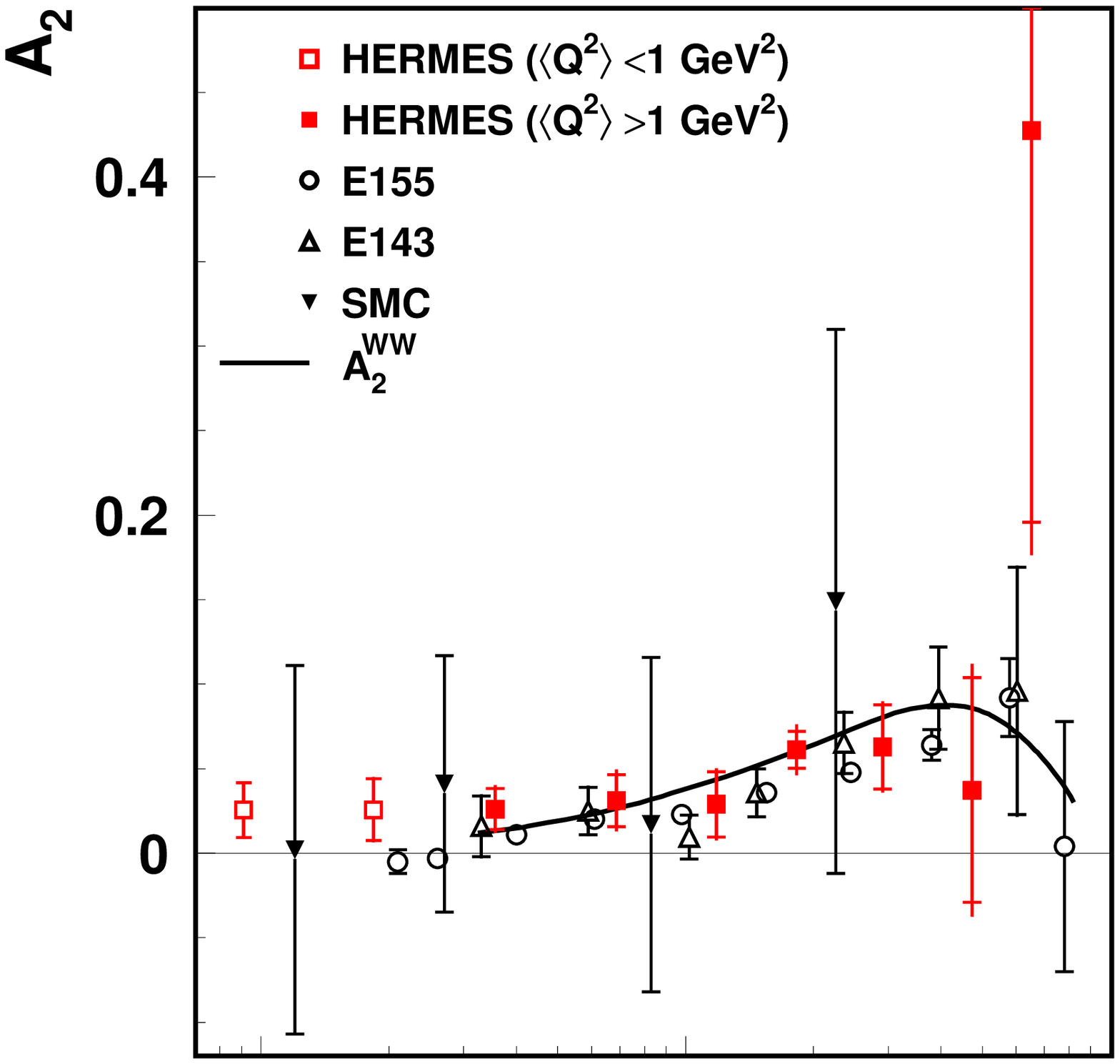}\\[-20.5mm]
\hspace*{-5mm}\includegraphics[width=10.1cm] {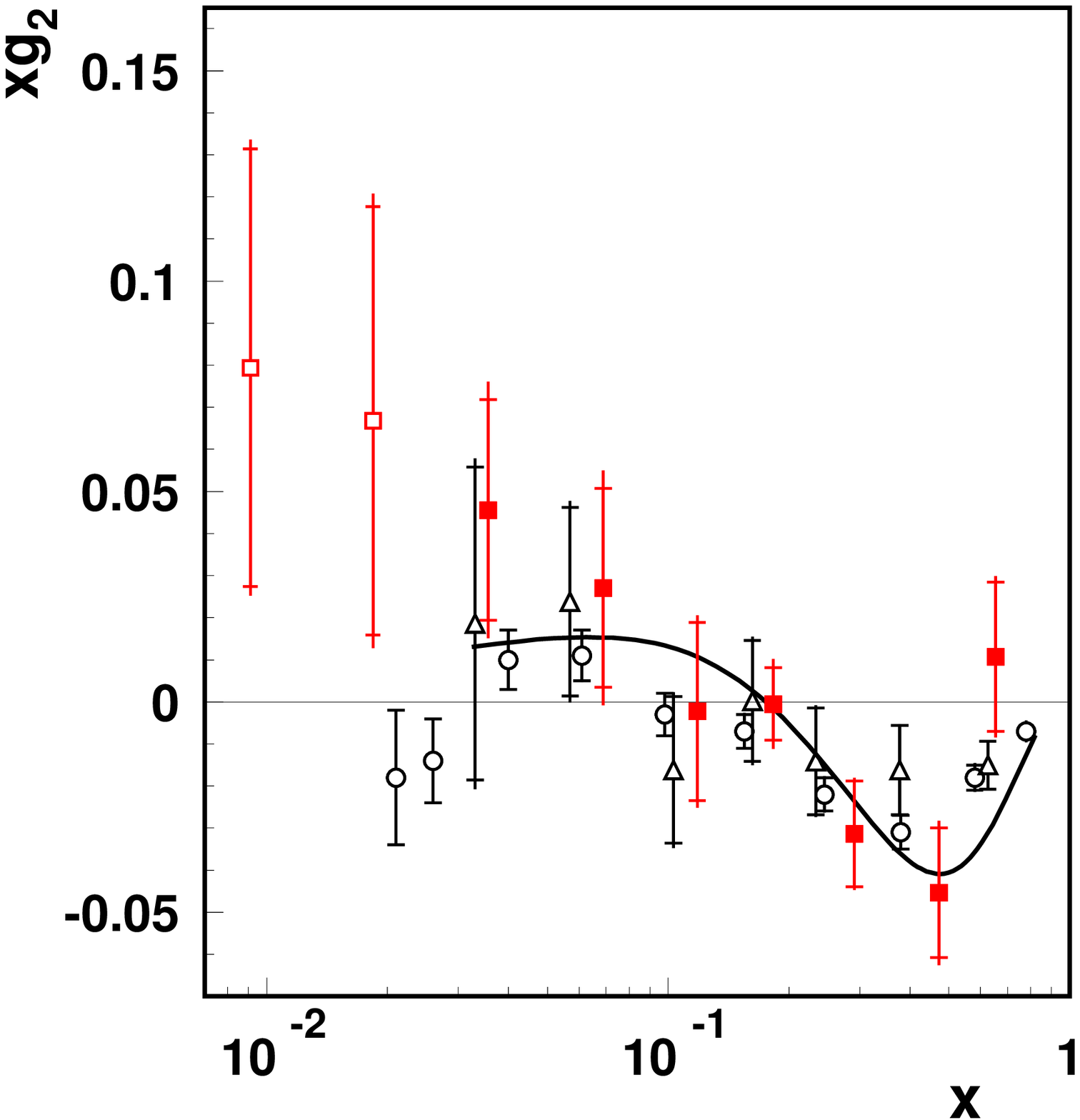}
\caption {
Upper panel: The virtual-photon asymmetry $A_2$ of the proton as a function of $x$.
Bottom panel: The spin-structure function $x g_2$ of the proton as a function of $x$.
HERMES data are shown together with data from the E155 \cite{e155}, E143 \cite{e143},
and SMC \cite{smc97} experiments. The total error bars for the HERMES, E155, and E143 experiments 
represent the quadratic sum of the statistical and systematic uncertainties. 
The statistical uncertainties are indicated by the inner error bars. 
The error bars for the SMC experiment represent the statistical uncertainties only.
The solid curve corresponds to the  
Wandzura--Wilczek relation (\ref{eq:g2ww}) evaluated at the average $Q^2$ values of HERMES
at each value of $x$. For the HERMES data, the closed (open) symbols represent data
with $\langle Q^2 \rangle \, > \, 1$~GeV$^2$ ($\langle Q^2 \rangle \, < \, 1$~GeV$^2$)}
\label{fig:figg2}
\end{figure}

The results for the virtual-photon asymmetry $A_2$ and the spin-structure 
function $x g_2$ as a function of $x$ are presented in Fig.~\ref{fig:figg2} together with
data from the experiments E155 \cite{e155}, E143 \cite{e143}, and
SMC \cite{smc97}. The HERMES data are shown for two regions of $Q^2$, 
$\langle Q^2 \rangle > 1$~GeV$^2$ (closed symbols) and 
$\langle Q^2 \rangle < 1$~GeV$^2$ (open symbols).
The experiments have only slightly different values
of average $Q^2$ for a particular value of $x$. The results are within their 
uncertainties in good agreement with each other. 
The solid curves represent
values of $A_2$ and  $x g_2$ evaluated with the Wandzura--Wilczek relation 
(\ref{eq:g2ww}) using the $g_1(x,Q^2)$ parameterization \cite{Ant00}. 
The values were calculated at the average $Q^2$ of HERMES at each value of $x$.
Within the uncertainties
the data satisfy the positivity bound \cite{positivity} for the asymmetry $A_2$,
$|A_2| \leq \sqrt{R (1 + A_1)/2} \, \simeq \, 0.4$, for all values of $x$ in the kinematic
range of the HERMES experiment.

The Burkhardt--Cottingham integral (\ref{eq:bcsumrule}) was evaluated in the measured region 
of $0.023 \leq x < 0.9$ at $Q^2 = 5$~GeV$^2$, resulting in
$ \int_{0.023}^{0.9} g_2(x, Q^2) \, dx = 0.006 \pm 0.024 \pm 0.017$. This result is to be 
compared with the combined result from experiments E143 and E155 \cite{e155} in the region
$0.02 \leq x < 0.8$: $ \int_{0.02}^{0.8} g_2(x, Q^2) \, dx = -0.042 \pm 0.008$.

Using the results measured by HERMES for the function $g_2$, 
the twist-3 matrix element $d_2$ given by (\ref{eq:d2mom})
was evaluated. For the unmeasured region $0.9 < x \leq 1$, the ansatz
$g_2(x) \propto (1-x)^3$ was assumed. The uncertainty in the extrapolated
contribution was taken to be equal to the contribution itself. 
The contribution from the region $x < 0.023$
was neglected because of the $x^2$ suppression factor. 
The result is $d_2 = 0.0148 \pm 0.0096(stat) \pm 0.0048(syst)$.
This is to be compared with the combined result from experiments E143 and E155 \cite{e155}: 
$d_2 = 0.0032 \pm 0.0017$.

In conclusion, HERMES measured the spin-structure function $g_2$ 
and the virtual-photon asymmetry $A_2$ of the proton in the kinematic range
$0.004 < x < 0.9$ and 0.18 GeV$^2$ $<$ $Q^2$ $<$ 20 GeV$^2$.
For the covered $x$-range the measured integral of $g_2(x)$  converges to
the null result of the Burkhardt--Cottingham sum rule.  
The $x^2$ moment of the twist-3 contribution to $g_2(x)$
is found to be compatible with zero, in agreement with expectations on its smallness 
from lattice calculations.
The results on $A_2$ and $g_2$ are overall in good agreement with measurements of SMC
at CERN, and E143 and E155 at SLAC, but they are not statistically precise enough 
to detect a deviation of $g_2$ from its Wandzura--Wilczek part, as seen by 
the SLAC experiments.

\begin{acknowledgement}
We gratefully acknowledge the \desy\ management for its support and the staff
at \desy\ and the collaborating institutions for their significant effort.
This work was supported by 
the Ministry of Economy and the Ministry of Education and Science of Armenia;
the FWO-Flanders and IWT, Belgium;
the Natural Sciences and Engineering Research Council of Canada;
the National Natural Science Foundation of China;
the Alexander von Humboldt Stiftung,
the German Bundesministerium f\"ur Bildung und Forschung (BMBF), and
the Deutsche Forschungsgemeinschaft (DFG);
the Italian Istituto Nazionale di Fisica Nucleare (INFN);
the MEXT, JSPS, and G-COE of Japan;
the Dutch Foundation for Fundamenteel Onderzoek der Materie (FOM);
the Russian Academy of Science and the Russian Federal Agency for 
Science and Innovations;
the U.K.~Engineering and Physical Sciences Research Council, 
the Science and Technology Facilities Council,
and the Scottish Universities Physics Alliance;
the U.S.~Department of Energy (DOE) and the National Science Foundation (NSF);
the Basque Foundation for Science (IKERBASQUE) and the UPV/EHU under program UFI 11/55;
and the European Community Research Infrastructure Integrating Activity
under the FP7 "Study of strongly interacting matter (HadronPhysics2, Grant
Agreement number 227431)".
\end{acknowledgement}

\begin{table*}
\caption{The spin-structure function $xg_2(x, Q^2)$ and 
the virtual-photon asymmetry $A_2(x, Q^2)$ of the proton
in bins of $(x, Q^2)$, see text for details. 
Statistical and systematic uncertainties are presented separately }
\label{tab:bigtab}
\begin{center}
\begin{tabular}{ccc|ccc|ccc}
\hline
\hline
bin & $\langle x\rangle$ &   $\langle Q^2\rangle,\mathrm{GeV^2}$  & $xg_2$   &     $\pm$ stat. & $\pm$ syst. & $A_2$  &    $\pm$ stat.   &   $\pm$ syst.\\
\hline
 1 &   0.009 &    0.38 &     0.0799 &     0.0521 &     0.0182 &     0.0257 &     0.0163 &     0.0057 \\
 2 &   0.018 &    0.68 &     0.0699 &     0.0513 &     0.0111 &     0.0269 &     0.0183 &     0.0040 \\
 3 &   0.033 &    0.89 &     0.0450 &     0.0326 &     0.0215 &     0.0278 &     0.0165 &     0.0109 \\
 4 &   0.039 &    1.37 &    -0.0047 &     0.0652 &     0.0080 &     0.0033 &     0.0275 &     0.0035 \\
 5 &   0.044 &    1.80 &     0.3489 &     0.1279 &     0.0612 &     0.1440 &     0.0507 &     0.0243 \\
 6 &   0.067 &    1.09 &     0.0044 &     0.0421 &     0.0097 &     0.0190 &     0.0346 &     0.0085 \\
 7 &   0.069 &    1.88 &     0.0473 &     0.0357 &     0.0062 &     0.0402 &     0.0210 &     0.0041 \\
 8 &   0.076 &    2.79 &     0.0202 &     0.0674 &     0.0323 &     0.0225 &     0.0342 &     0.0164 \\
 9 &   0.116 &    1.30 &    -0.0094 &     0.0506 &     0.0081 &     0.0266 &     0.0603 &     0.0111 \\
10 &   0.118 &    2.44 &     0.0356 &     0.0301 &     0.0099 &     0.0584 &     0.0251 &     0.0090 \\
11 &   0.124 &    4.04 &    -0.0571 &     0.0466 &     0.0149 &    -0.0137 &     0.0311 &     0.0102 \\
12 &   0.182 &    1.51 &    -0.0758 &     0.0642 &     0.0230 &    -0.0466 &     0.1055 &     0.0389 \\
13 &   0.183 &    3.01 &     0.0121 &     0.0324 &     0.0038 &     0.0707 &     0.0375 &     0.0074 \\
14 &   0.187 &    5.42 &    -0.0334 &     0.0440 &     0.0041 &     0.0143 &     0.0392 &     0.0052 \\
15 &   0.282 &    1.95 &     0.0071 &     0.0396 &     0.0063 &     0.1675 &     0.0925 &     0.0167 \\
16 &   0.298 &    3.99 &    -0.0242 &     0.0195 &     0.0055 &     0.0718 &     0.0363 &     0.0117 \\
17 &   0.311 &    7.58 &    -0.0571 &     0.0283 &     0.0105 &     0.0039 &     0.0437 &     0.0166 \\
18 &   0.458 &    2.83 &    -0.0613 &     0.0582 &     0.0129 &     0.0064 &     0.2616 &     0.0598 \\
19 &   0.482 &    4.31 &    -0.0987 &     0.0370 &     0.0104 &    -0.2064 &     0.1704 &     0.0500 \\
20 &   0.484 &    7.57 &    -0.0362 &     0.0183 &     0.0045 &     0.0421 &     0.0744 &     0.0206 \\
21 &   0.630 &    4.76 &     0.2413 &     0.1194 &     0.0534 &     3.0231 &     1.3295 &     0.5969 \\
22 &   0.658 &    6.79 &    -0.0129 &     0.0320 &     0.0081 &     0.1197 &     0.4350 &     0.1115 \\
23 &   0.678 &   10.35 &     0.0076 &     0.0160 &     0.0025 &     0.3672 &     0.2551 &     0.0419 \\
\hline
\hline
\end{tabular}
\end{center}

\end{table*}
\begin{table*}
\caption{
The spin-structure function $xg_2$ and 
the virtual-photon asymmetry $A_2$ of the proton after evolving to common $Q^2$
and averaging over in each $x$-bin (see text for details). 
Statistical and systematic uncertainties are presented separately
}
\label{tab:tab1}
\begin{center}
\begin{tabular}{cccc|ccc|ccc}
\hline
\hline
bin& $x$ range & $\langle x\rangle$ & {\footnotesize $\langle Q^2\rangle$,GeV$^2$}
 &
$xg_2$ & $\pm$stat & $\pm$syst & $A_2$ & $\pm$stat & $\pm$syst\\
\hline
 1 & 0.004~-~0.014 & ~0.009 &  0.38 &~ 0.0794~& 0.0520 & 0.0153 ~&~ 0.0256~& 0.0162 & 0.0049 \\
 2 & 0.014~-~0.023 & ~0.018 &  0.68 &~ 0.0668~& 0.0509 & 0.0181 ~&~ 0.0258~& 0.0182 & 0.0065 \\
 3 & 0.023~-~0.050 & ~0.036 &  1.08 &~ 0.0456~& 0.0262 & 0.0157 ~&~ 0.0261~& 0.0121 & 0.0074 \\
 4 & 0.050~-~0.090 & ~0.069 &  1.59 &~ 0.0271~& 0.0236 & 0.0150 ~&~ 0.0312~& 0.0154 & 0.0100 \\
 5 & 0.090~-~0.150 & ~0.118 &  2.07 &~-0.0023~& 0.0212 & 0.0085 ~&~ 0.0289~& 0.0194 & 0.0088 \\
 6 & 0.150~-~0.220 & ~0.183 &  2.51 &~-0.0005~& 0.0086 & 0.0063 ~&~ 0.0612~& 0.0109 & 0.0105 \\
 7 & 0.220~-~0.400 & ~0.291 &  3.23 &~-0.0314~& 0.0126 & 0.0043 ~&~ 0.0629~& 0.0248 & 0.0104 \\
 8 & 0.400~-~0.600 & ~0.473 &  4.62 &~-0.0454~& 0.0154 & 0.0075 ~&~ 0.0373~& 0.0665 & 0.0345 \\
 9 & 0.600~-~0.900 & ~0.654 &  7.06 &~ 0.0107~& 0.0177 & 0.0073 ~&~ 0.4275~& 0.2316 & 0.0970 \\
\hline
\hline
\end{tabular}
\end{center}

\end{table*}
\begin{table*}
\caption{Correlation matrix for $x g_2$ in 9 $x$-bins (as in Table~\ref{tab:tab1}) }
\label{tab:tab2}
\begin{center}
\begin{tabular}{l|ccccccccc}
\hline
\hline
~  & 1       & 2       &  3      & 4       & 5       & 6       & 7       & 8       & 9 \\
\hline
 1 &  1.0000 & -0.1281 & -0.0038 & -0.0033 & -0.0017 &  0.0005 &  0.0000 &  0.0000 &  0.0000 \\
 2 & -0.1281 &  1.0000 & -0.1584 & -0.0083 & -0.0007 &  0.0000 &  0.0000 &  0.0001 &  0.0000 \\
 3 & -0.0038 & -0.1584 &  1.0000 & -0.1951 & -0.0281 &  0.0077 & -0.0016 &  0.0002 &  0.0000 \\
 4 & -0.0033 & -0.0083 & -0.1951 &  1.0000 & -0.2885 &  0.0312 & -0.0107 &  0.0013 & -0.0005 \\
 5 & -0.0017 & -0.0007 & -0.0281 & -0.2885 &  1.0000 & -0.0102 & -0.0654 &  0.0067 & -0.0018 \\
 6 &  0.0005 &  0.0000 &  0.0077 &  0.0312 & -0.0102 &  1.0000 & -0.1829 &  0.0143 & -0.0055 \\
 7 &  0.0000 &  0.0000 & -0.0016 & -0.0107 & -0.0654 & -0.1829 &  1.0000 & -0.3539 &  0.0926 \\
 8 &  0.0000 &  0.0001 &  0.0002 &  0.0013 &  0.0067 &  0.0143 & -0.3539 &  1.0000 & -0.3947 \\
 9 &  0.0000 &  0.0000 &  0.0000 & -0.0005 & -0.0018 & -0.0055 &  0.0926 & -0.3947 &  1.0000 \\
\hline
\hline
\end{tabular}
\end{center}

\end{table*}

\begin{thebibliography}{99}
\bibitem{anselmino}
M. Anselmino, A. Efremov, E. Leader, \PR{261}{1}{1995}.
\bibitem{shuryak}
E.V. Shuryak and A. I. Vainshtein, \NPB{201}{141}{1982}.
\bibitem{Jaf90}
R.L. Jaffe, Comments Nucl. Part. Phys. {\bf 19}, 239 (1990).
\bibitem{jaffe-ji91}
R.L. Jaffe and X. Ji, \PRD{43}{724}{1991}.
\bibitem{Wan77}
S. Wandzura and F. Wilczek, \PLB{72}{195}{1977}.
\bibitem{Gockeler}
M. G\"ockeler \etal, \PRD{63}{074506}{2001}.
\bibitem{Gockeler2}
M. G\"ockeler \etal, \PRD{72}{054507}{2005}.
\bibitem{sivers}
D.W. Sivers, \PRD{41}{83}{1990}.
\bibitem{burkardt}
M. Burkardt, e-Print: arXiv:0810.3589
\bibitem{hermessivers}
HERMES Coll.,  A. Airapetian \etal, \PRL{103}{152002}{2009}.
\bibitem{bcsumrule}
H. Burkhardt and W.N. Cottingham, \ANN{56}{453}{1970}.
\bibitem{smc97}
SMC Coll., D. Adams \etal, \PRD{56}{5330}{1997}.
\bibitem{e143}
E143 Coll., K. Abe \etal, \PRD{58}{112003}{1998}.  
\bibitem{e155}
E155 Coll., P. L. Anthony \etal, \PLB{553}{18}{2003}. 
\bibitem{HERM07}
HERMES Coll., A. Airapetian \etal, \PRD{75}{012007}{2007}.
\bibitem{hermes}
HERMES Coll., K. Ackerstaff \etal, \NIMA{417}{230}{1998}.
\bibitem{target}
HERMES Coll., A. Airapetian \etal, \NIMA{540}{68}{2005}.
\bibitem{abs2003}
C. Baumgarten \etal ~~(HERMES Target Group), \NIMA{496}{606}{2003}.
\bibitem{polarmeas}
C. Baumgarten \etal ~~(HERMES Target Group), \NIMA{508}{633}{2003}.
\bibitem{sok}
A.A. Sokolov, I.M. Ternov, Sov. Phys. Doklady {\bf 8}, 1203 (1964).
\bibitem{buon}
J. Buon and K. Steffen, \NIMA{245}{248}{1986}.
\bibitem{barber}
D.P.~Barber \etal, \PLB{343}{436}{1995}.
\bibitem{tpol}
D.P.~Barber \etal, \NIMA{329}{79}{1993}.
\bibitem{lpol}
HERMES Coll., M. Beckmann \etal, \NIMA{479}{334}{2002}.
\bibitem{lum}
HERMES Coll., T. Benisch \etal, \NIMA{471}{314}{2001}.
\bibitem{Aku98}
I. Akushevich, H. B\"ottcher and D. Ryckbosch (1998), hep-ph/9906408.
\bibitem{HERM05}
HERMES Coll., A. Airapetian \etal, \PRD{71}{012003}{2005}.
\bibitem{HERMb1} 
HERMES Coll., A. Airapetian \etal, \PRL{95}{242001}{2005}.
\bibitem{Ant00}
E155 Coll., P.L. Anthony \etal, \PLB{493}{19}{2000}. 
\bibitem{Gab07}
HERMES Coll., A. Airapetian \etal, \JHEP{0511}{126}{2011}.
\bibitem{Abe99}
E143 Coll., K. Abe \etal, \PLB{452}{194}{1999}.  
\bibitem{positivity}
J. Soffer and O.V. Teryaev, \PLB{490}{106}{2000}.
\end{thebibliography}
\end{document}